\documentclass[pra,twocolumn,amsfonts,superscriptaddress,a4paper,showpacs,showkeys]{revtex4}
\usepackage{graphicx} 
\newcommand{\ga}{\alpha}
\newcommand{\gb}{\beta}

\newcommand {\sg} {\sigma}

\newcommand {\gth} {\theta}

\begin{document}

\title{ Short-time dynamics of random-bond Potts ferromagnet \\
with continuous self-dual quenched disorders\footnote{Supported by the 
NNSF of China, 19975041 and 10074055, 
and by the Deutsche Forschungsgemeinschaft, DFG Schu 95/9-3.} }

\author{Z.Q. Pan}
\affiliation{Physics Department, Zhejiang University, Hangzhou 310027, P.R. China}
\author{H.P. Ying$^{1,}$}
\email{hpying@zimp.zju.edu.cn}
\affiliation{Universit{\"a}t Siegen, D-57068 Siegen, Germany}
\author{D.W. Gu}
\affiliation{Hangzhou Science and Technology University for Adult, Hangzhou 
310012, PR China \\~\\}

\begin{abstract}
We present Monte Carlo simulation results of random-bond Potts ferromagnet
with the {\it Olson-Young} self-dual distribution of quenched disorders in
two-dimensions. By exploring the short-time scaling dynamics, we find
universal power-law critical behavior of the magnetization and Binder cumulant
at the critical point, and thus obtain estimates of the dynamic exponent $z$
and magnetic exponent $\eta$, as well as the exponent $\theta$.
Our special attention is paid to the dynamic process for the $q$=8 Potts model.
\end{abstract}

\pacs{75.40.Mg, 64.60.Fr, 64.60.Ht\\
Keywords: Short-time dynamics, Potts model, Critical exponents, Universality}

\maketitle

\section{\bf Introduction}

Ferromagnetic systems with quenched randomness are studied intensively
for their critical properties affected by an addition of
disorders \cite{Cardy96,Jaco98,Folk00}.
The Harris criterion was claimed 27 years ago that if the specific heat
critical exponent $\alpha$ of the pure system is positive, then quenched
disorder is a relevant perturbation at the {\it second-order} critical point
and it causes changes in critical exponents \cite{Harris}.
Then, following the work of Imry and Wortis \cite{Imry79} who
argued that quenched disorder could smooth of a {\it first-order} phase
transition and thus produce a new {\it second-order}
phase transition, the introduction of randomness to pure systems originally
undergoing a first-order transition has been comprehensively considered
\cite{Hui89,Aizen89,Cardy99}.
The theory was firstly numerically checked with the Monte Carlo (MC) method
by Chen, Ferrenberg and Landau (CFL) \cite{Chen92} who studied
the 8-state Potts model with a self-dual random-bond disorder
(see Eqs.(\ref{eq2}) and (\ref{eq3})).
Recently most of the work for such disordered systems is carried out by
intensive MC simulations \cite{Wise95,Kardar95,Kim96,Yasar98,Chate98,Olson99}
for examining how a phase transition is modified
by the quenched disorder coupling to the local energy density
for the $q$-state random-bond Potts ferromagnet (RBPF).

The $q$-state RBPF, an interesting framework to study the influence of
impurity on pure systems, is described by the Hamiltonian,
\begin{eqnarray}
-\beta H=\sum_{<i,j>} K_{ij} \delta_{\sg_i \sg_j}~,~~~~ K_{ij} > 0,~~~~~
\label{Ham1}
\end{eqnarray}
where $\beta=1/k_B T$ is the inverse temperature, the spin $\sg$ can take the
values 1,$\cdots,q$, $\delta$ stands for the Kronecker {\it delta} function,
and the nearest-neighbor interactions $<i,j>$ are considered. Usually
the dimensionless couplings $K_{ij}$ are selected from two positive
(ferromagnetic) values $K_1$ and $K_2=rK_1$, with a strong
to weak coupling ratio $r=K_2/K_1$, called a {\it disorder amplitude},
according to a bimodal distribution,
\begin{eqnarray}
P(K)=p\delta (K -K_1) + (1-p)\delta (K -K_2)~.
\label{eq2}
\end{eqnarray}
With $p=0.5$, the system is {\it self-dual} and the exact
critical point can be determined by \cite{Kin81},
\begin{eqnarray}
( e^{K_{c}} - 1)(e^{K'_{c}} - 1)= q~,
\label{eq3}
\end{eqnarray}
where $K_{c}$ and $K'_{c}$ are the corresponding critical values of $K_1$ and
$K_2$, respectively. The value $r=1$ corresponds to the pure
case where the critical point is located at $K_c=\mbox{log}(1+\sqrt {q~})$
and the phase transitions are first-order for $q > 4$. As $\ga>0$ for all
$q>2$ the randomness acts as a relevant perturbation and, for $q>4$,
it even changes the nature of the transition from first to second order.

Up to present, the MC studies are usually carried out in {\it equilibrium}
and concentrated on disorder amplitudes in the
range of $r=2$-20 adapted to numerical analysis. They give good estimates
of the disordered fixed point exponents \cite{Cardy99,Picco98}.
In Table 1 we present the magnetic scaling index $\eta=2\gb/\nu$ of the 2D
8-state RBPF obtained by different groups
\cite{Cardy97,Chate98a,Chate99,Picco98,Ying00}.
In a recent work, Olson and Young (OY) \cite{Olson99} used their special
{\it self-dual} disorder distribution and performed the
MC study of {\it multiscaling} properties of the correlation functions for
several values of $q$. Their results for the magnetic exponent are
very interesting to examine the universality of the RBPF in the crossover
regime from the pure fixed point to a percolation-like limit \cite{Picco98}.
Cardy and Jacobsen \cite{Cardy97} studied the RBPF based on the connectivity
transfer matrix (TM) formalism, and their estimates of the critical
exponents lead to a continuous variation of $\eta$ with $q$, which is in
sharp disagreement with CFL's results \cite{Chen92}.

\begin{table} \begin{center}
\caption {
Magnetic exponent $\eta$ estimated by different
groups for the 2D 8-state RBPF. }
\vskip 0.5cm
\begin{tabular}{c  c  c  c } \hline\hline
      Authors      & $r$   &  $\eta$ & Technique \\ \hline
    CFL\cite{Chen92}               & ~2,10~ & 0.236(4) & MC \\ \hline
Olson and Young\cite{Olson99}      &$P_X(x)$& 0.319(7) & MC  \\ \hline
Cardy and Jacobsen\cite{Cardy97}   & ~~2~~~ & 0.284(4) & TM \\ \hline
Chatelain and Berche\cite{Chate98a}& ~~10~~ & 0.304(6) & MC  \\ \hline
Chatelain and Berche\cite{Chate99} & ~~10~~ & 0.301(1) & TM  \\ \hline
Picco\cite{Picco98}                & ~~10~~ & 0.306(2) & MC  \\ \hline
Ying and Harada\cite{Ying00}       & ~~10~~ & 0.302(5) & STD \\ \hline\hline
\end{tabular}
\end{center} \end{table}

In this paper, we present a MC study to verify the dynamic scaling of the
RBPF and estimate the critical exponents in the short-time dynamics
(STD) \cite{Janss89,Zheng98} for the OY's
{\it continuous} distribution of the disorder interactions. It is a
self-dual scheme to introduce quenched randomness and defined by
the following probability
\begin{eqnarray}
P_X(x) =\frac{2\sqrt{q}}{\pi(1-x)^2+qx^2},~~~~~
\label{OY}
\end{eqnarray}
for the Boltzmann factor $x\equiv e^{-K}$. It is obvious that the model
has strong disorder as the distribution function is nonzero everywhere at
$x\in(0,1)$ ($K\rightarrow\infty$, $K\rightarrow 0$) and has a finite
weight at its two limit points $x=0$ and $x=1$.
Another advantage of this distribution is that it is easy to generate
random numbers with the probability $P_X(x)$ through
\begin{eqnarray}
x = \frac{1} {1+\sqrt{q}\tan (\pi R_x/2)},~~~~~~~~
\label{OYa}
\end{eqnarray}
where $R_x$ is a random number with uniform distribution between 0 and 1.

In particular, we will investigate the critical behavior affected by
introduction of such continuous quenched randomness
to clarify the universality class of the RBPF in the crossover regime.
In simulations we mostly choose the model with $q=8$ which is known to have
a strong first-order phase transition without disorders. We hope to confirm
the new second-order phase transition induced by the random disorder in
order to show the influence of quenched impurities on the first-order
systems and to check whether an Ising-like universality would be satisfied
from the STD approach.

\section{\bf Short-time Dynamics}

For a long time, it was believed that universality and scaling relations
can be found only in equilibrium or in the {\it long-time} regime. In Refs.
\cite{Janss89,Zheng98},  however it was discovered that for a $O(N)$
vector magnetic system in states with a very high temperature
$T \gg T_c$, when it is suddenly quenched to the critical temperature $T_c$,
and evolves according to a dynamics of model A, a universal dynamic scaling 
behavior emerges already within the short-time regime,
\begin{equation}  \label{eq4}
M^{(k)}(t,\tau,L,m_0)=b^{-k\gb/\nu}M^{(k)}(b^{-z}t,b^{1/\nu}\tau,
b^{-1}L,b^{x_0}m_0), 
\end{equation}
where $M^{(k)}$ is the $k$th moment of magnetization, $t, \tau=(T-T_c)/T_c,
L$ and $b$ are time, reduced temperature, lattice size and
scaling factor, respectively. $\gb$ and $\nu$ are the well known static
critical exponents. The $x_0$, a {\it new independent} exponent, is the
scaling dimension of initial magnetization $m_0$.  This dynamic scaling
form is generalized from finite size scaling in the equilibrium stages.
The dynamic MC simulations have been successfully performed in such
non-equilibrium processes to estimate the critical point and the critical
exponents ($K_c$,$z$,$\gb$,$\nu$,$\gth$) \cite{Li95,Zheng99,Ying98,Ying01}.

The MC simulation begins with a study of the time evolution in the
initial stage of the dynamic relaxation starting at very high temperature
and small magnetization ($m_0\sim 0$). For a sufficiently large lattice
($L\rightarrow\infty$), from Eq.(\ref{eq4}) by setting $\tau=0$ and
$b = t^{1/z}$, it is easy to derive that
\begin{eqnarray}
 M^{(k)}(t,m_{0})=t^{-k\beta/\nu z}M^{(k)}(1,t^{x_{0}/z}m_{0}).~~~~
\label{eq5}
\end{eqnarray}
When $k=1$ we get the most important scaling relation on which
our measurements of the critical exponent $\gth$ are based,
\begin{eqnarray}
M(t) \sim m_0 t^\gth,~~~~\gth = (x_0 - \beta/\nu)/z.~~
\label{theta}
\end{eqnarray}
In most cases the magnetization undergoes an initial increase at the critical
point $K_c$ after a microscopic time $t_{mic}$.

In our MC sweeps, the time evolution of $M(t)$ is calculated through
the definition
\begin{eqnarray}
 M(t)&=&\frac{1}{N}\left [\left < \frac{q M_O -N}{q-1} \right >\right ].~~~~~~~~
\label{eq6}
\end{eqnarray}
Here $M_O =\mbox{max}(M_1, M_2, \cdots, M_q)$ with $M_i$ being the
number of spins in the $i$th state among $q$ states.
$N=L^2$ is the number of spins on the suqare lattice.
We use $L$ up to 128. $<\cdots>$ denotes thermal averages
over independent initial states and random number sequences, and
$[\cdots]$ the disorder averages over quenched randomness distributions.
The unit of time $t$ is defined as a MC sweep over all spins on the lattice.

The susceptibility plays an important role in the equilibrium. Its finite
size behavior has been often used to determine the critical temperature
and the critical exponents $\gamma/\nu$ and $\beta/\nu$ \cite{Chen92}.
For the STD approach, the time-dependent susceptibility (the second
moment of the magnetization) is also interesting and important.
For the random-bond Potts model, it is defined as
\begin{eqnarray}
M^{(2)}(t)&=&\frac{1}{N}\left [\left (<M^2(t)>-<M(t)>^2\right )\right ].~~~~
\label{eq_eta}
\end{eqnarray}
One studies its scaling behavior from $m_0 =0$ in short-time relaxation.
Because the spatial correlation length in the beginning of
the evolution is small compared with the lattice size $L^d$
the second moment behaves as $M^{(2)}(t,L)\sim L^{-d}$. Then the
finite  size scaling Eq.(\ref{eq4}) induces a power-law
behavior at the critical temperature,
\begin{eqnarray}
 M^{(2)}(t) \sim t^{y},~~ y=(d-\eta)/z.~~
\label{index_y}
\end{eqnarray}

In the above considerations the dynamic relaxation was assumed to start from
disordered states with $m_0$ small enough. Another interesting and important
process is the dynamic relaxation from a completely ordered state. The initial
magnetization is taken exactly at its other fixed point $m_0=1$, where
a critical scaling form
\begin{eqnarray}
M^{(k)}(t,\tau,L)=b^{-k\beta/\nu}M^{(k)}(b^{-z}t,b^{1/\nu}\tau, b^{-1}L) ~~~~~
\label{eq8}
\end{eqnarray}
is expected \cite{Ito93,Zheng98}. This scaling form looks to be the same as
the dynamic scaling form in the long-time regime, however, it is now
assumed already valid in the macroscopic short-time regime.
For the magnetization itself, $b=t^{1/z}$ yields, for a sufficiently
large lattice and at the critical point ($\tau=0$),
a power-law decay behavior of
\begin{eqnarray}
M(t,\tau)=t^{-c_1} , ~~c_1 =\beta/\nu z. ~~
\label{index_c1}
\end{eqnarray}
The formula can be used to calculate
the critical exponents $\eta=2\beta/\nu$ or $z$.

We further observe an evolution of the Binder cumulant $U(t, L) =
M^{(2)}(t,L)/M^2(t,L) - 1$.
A similar power-law behavior at criticality induced from the
Eq.(\ref{eq8}) shows that
\begin{eqnarray}
 U(t, L) \sim t^{c_u}~, ~~~c_u = d/z ~~~~~~~~~~~~~
\label{index_cu}
\end{eqnarray}
on large enough lattices.
Here, unlike the relaxation from the disordered state, the
fluctuations caused by the initial configurations are much smaller.
In practical simulations, these measurements of the critical exponents
and critical temperature are better in quality than those from the
dynamical relaxation starting from disordered states.

\section{\bf MC Results}

In simulations, the heat-bath algorithm is used for MC updating, and
up to 500 samples are chosen as disorder averages and 200-500 initial
configurations and/or random number sequences are taken as thermal averages
for each disorder realization (so total MC averages are over 100,000).
Statistical errors are simply estimated by performing three groups of
averages with different random number sequences as well as independent
initial states. We first focus our attention on evolutions of the
magnetization both from the initial states with small magnetization
($m_0\sim 0$) and complete ordered states ($m_0=1$).

\begin{figure}[htbp!]\centering
\caption{$M(t)$ versus MC time plotted on a log-log scale and carried out on a
$64^2$ lattice from the initial states $m_0\sim 0$ at and near the
critical point, $K_c$ and $K_c^{(\pm)}=(1.00\pm 0.01)K_c$.
  }
\includegraphics*[width=7.30cm]{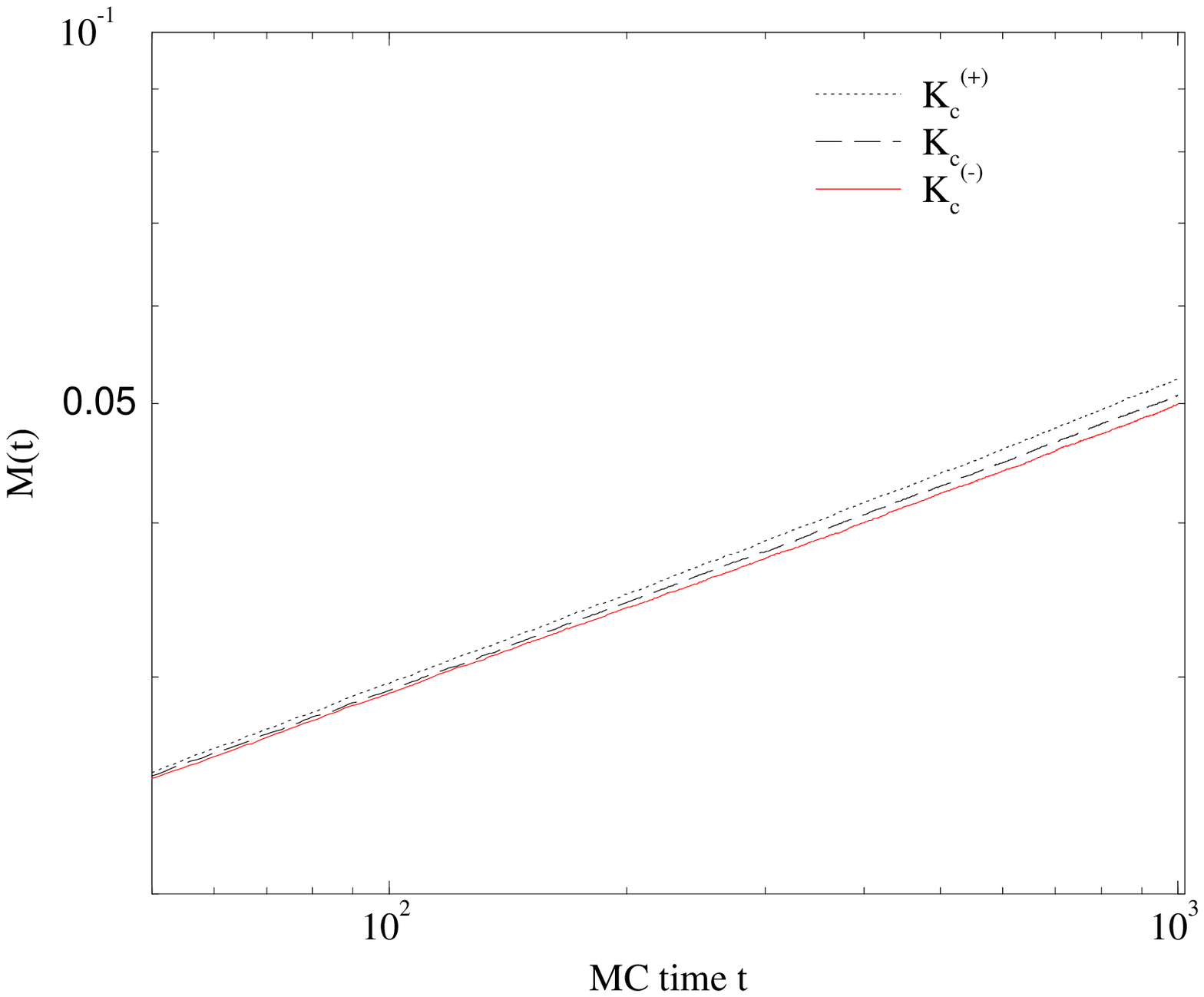}
\label{fig1}
\vspace{0.36cm}
\caption{$U(t)$ versus MC time plotted on a log-log scale and carried out on a
$64^2$ lattice from the initial state $m_0=1$ at and near the
critical point, $K_c$ and $K_c^{(\pm)}=(1.00\pm 0.01)K_c$.
  }
\includegraphics*[width=7.30cm]{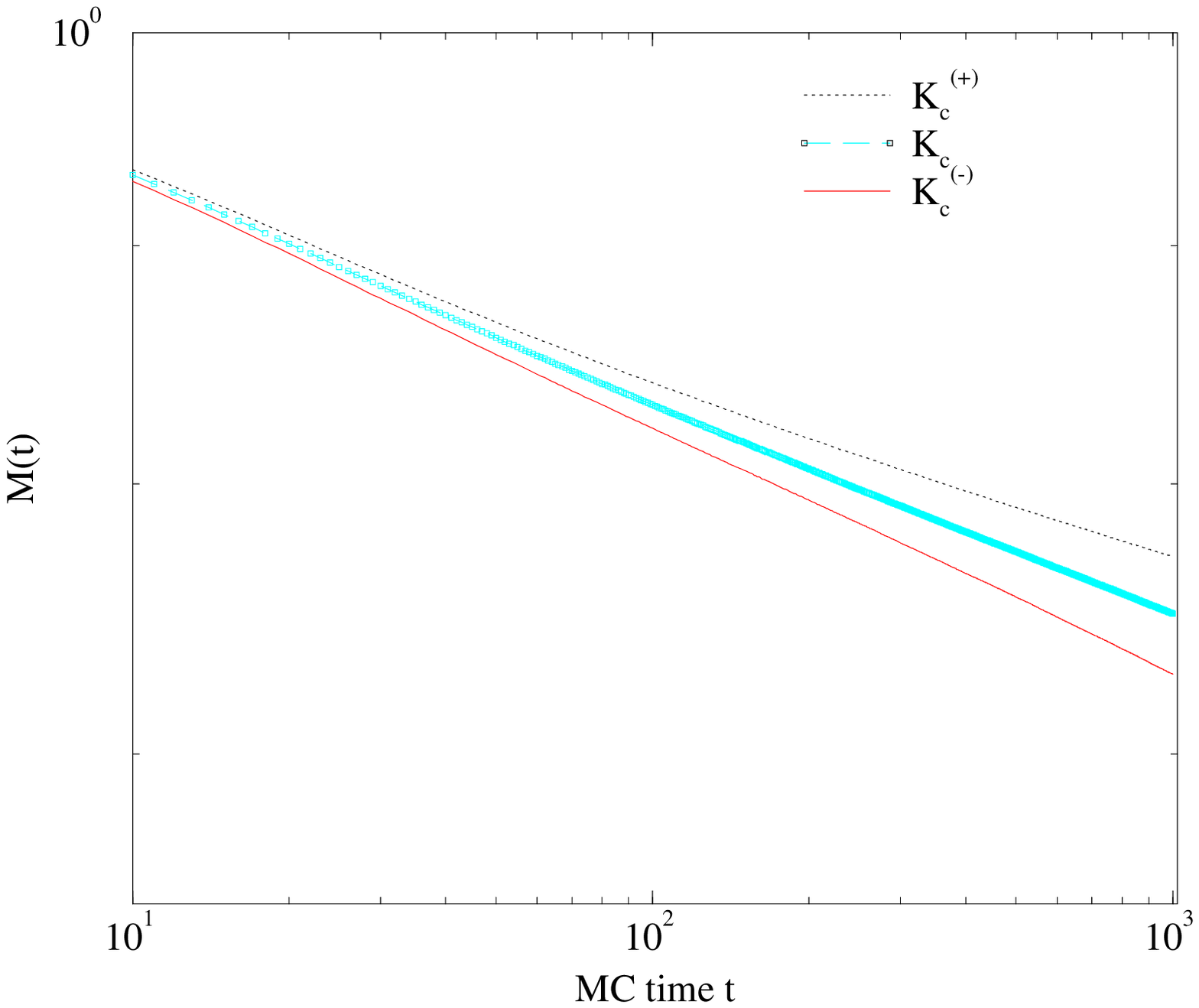}
\label{fig2}
\end{figure}

At the critical point ($\tau =0$), achieved by generating the Boltzmann
factors according to the distribution of Eq.({\ref{OYa}), the curves
$M(t)$ should show power-law behavior as described by Eq.(\ref{theta})
or Eq.(\ref{index_c1}) dependent on the initial states. They, however,
should deviate from the power-law scaling behavior for a small but nonzero
$\tau$ characterized by $M(t,\tau)=m_0 t^\gth M(t^{1/\nu z}\tau)$ or
$M(t,\tau)=t^{-c_1} F(t^{1/\nu z} \tau)$ with respect to the $m_0\sim 0$ or
$m_0=1$ initial states respectively.
Now a question arises how to introduce a distribution which
deviates from the critical point with distance $\tau$? We can first
generate the Boltzmann factors, $x\equiv e^{-K}$, from the self-dual
distribution $P_X(x)$ in Eq.(\ref{OY}), then modify them by the replacement
$x\rightarrow x^{1/(1+\tau)}$ away from criticality \cite{Olson99}.
We observe that the best power-law behavior appeared at the
same critical point for both the random and ordered initial states. This
fact indicates a second-order phase transition induced at $K_c$,
as argued by Sch\"ulke and Zheng \cite{Schu00} since
$K_{order} <K_c < K_{random}$ if a first-order phase transition happened.
We exhibit in Fig.1 and Fig.2 the double-logarithmic plots of magnetization
versus the MC time on a $L=64$ lattice to show this behavior obviously.
\begin{table} \begin{center}
\caption {
The values of indices $c_1,c_u$, $\theta$ and $y$ and estimated results $z$
and $\eta$ estimated from the $m_0=1$ state for the $q=8$ on lattice sizes $L^2$.
}
{\small
\begin{tabular}{c|cc|cc|cc} \hline\hline
  ~  &$m_0=1$   &  ~   &$m_0\sim0$& ~ &results&  ~  
\\ \hline
index~ & $c_1$    &  $c_u$  &  $\gth$ &   $y$   &  $z$    & $\eta$ 
\\ \hline
$L=32$& .0468(6) & .603(7) & .254(3) & .597(6) & 3.32(5) & .310(5)
\\
$L=64$& .0450(6) & .597(7) & .238(3) & .526(6) & 3.35(5) & .303(5)
\\
$L=96$& .0449(5) & .593(7) & .234(4) & .508(5) & 3.37(5) & .302(5)
\\
$L=128$& .0445(5) & .586(7) & .230(4) & .497(5) & 3.41(6) & .304(5)
\\ \hline\hline
\end{tabular}
   }
\end{center} \end{table}

Next, we carry out simulations at the exact critical point $K_c$ to
investigate the magnetization and Binder cumulant, starting from $m_0 =1$
ordered state. Such state is a ground state where all the
spins orient in the same direction (e.g., $\sg_i$ = 1, $i=1,N$).
In Figs.3 and 4, $M(t)$ and $U(t)$ are presented and values of the scaling
indices $c_1=\gb/\nu z$ and $c_u=d/z$ can be estimated from slopes of the
curves in $t=$(100,1000). The results are listed in Table 2.
Here we find that, unlike for relaxations from the disordered
state, the fluctuations caused by ordered initial configurations
are much smaller.
As a result, these measurements of the critical exponents based on
Eqs.(\ref{index_c1}) and ({\ref{index_cu}) are better in quality
than those from disordered states on Eqs.(\ref{theta}) and ({\ref{index_y}).
So we take only 200 MC configurations for the thermal averages in the MC
simulations. Now, from the results of $c_1$ and $c_u$, the critical exponents
$\eta$ and $z$ can be calculated based on Eqs.(\ref{index_c1}) and
(\ref{index_cu}), and their values are summarized in Table 2.
\begin{figure}[htbp!]\centering
\caption{Time evolution of magnetization for different lattice sizes
from the ordered state at the critical point, plotted on a
double-log scale. It shows the finite size effect on the $M(t)$.
    }
\includegraphics*[width=7.30cm]{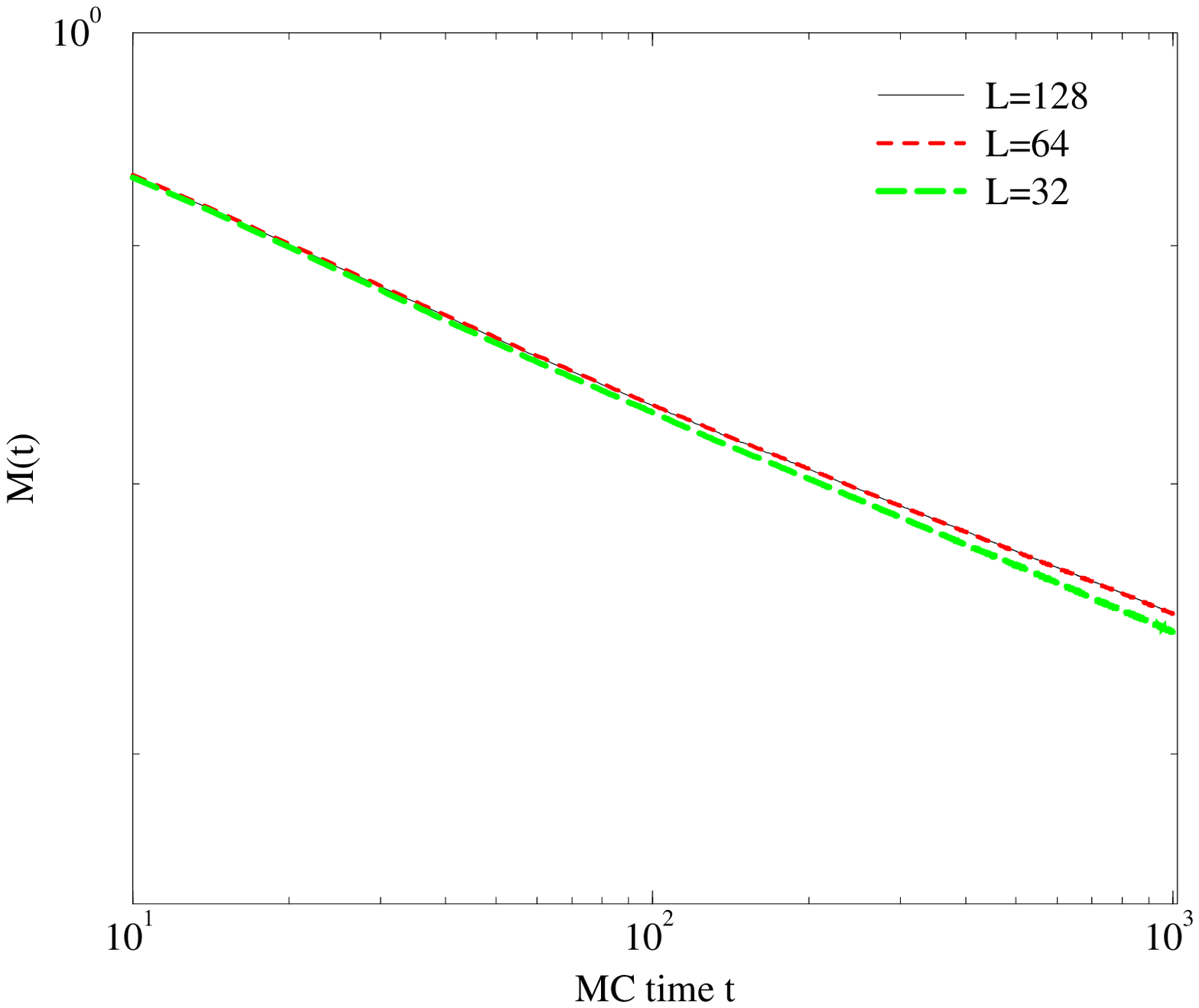}
\label{fig3}
\vspace{0.36cm}
\caption{Time evolution of Binder cumulant for different lattice sizes from
the ordered state at the critical point, plotted on a
double-log scale. The $t_{mic}$ is extended to about one hundred.
    }
\includegraphics*[width=7.30cm]{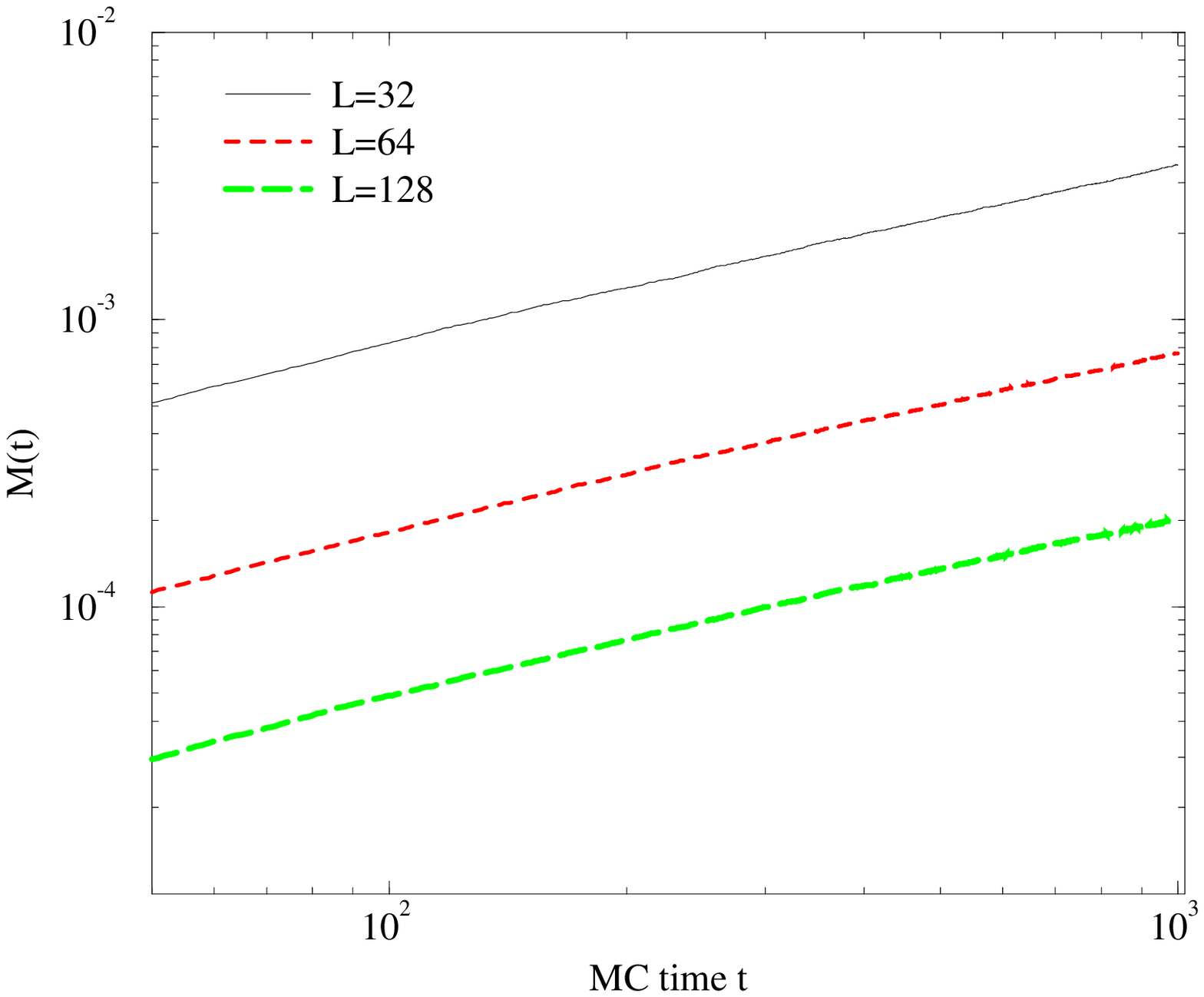}
\label{fig4}
\end{figure}
To estimate the critical exponent $\gth$, the random initial states are
prepared  with small magnetizations $m_0 = 0.04$ -- 0.01. The curves of
$M(t)$ for different $m_0$ on $128^2$ lattice with a double-logarithmic scale
are displayed in Fig.5 which exhibits the finite $m_0$ effect on $M(t)$
and a very nice power-law behavior at $m_0\rightarrow 0$. Thus the
exponent $\gth$ can be estimated from slopes of the curves in
$t$=(50,1000) by using Eq.(\ref{theta}) after an extrapolation to
$m_0 \rightarrow 0$. Its values are shown also in Table 2.

As mentioned above, for a determination of the dynamic exponent $z$ and the
magnetic exponent $\eta$ a dynamic process starting from the ordered} state
($m_0=1$) is more favorable since the thermal fluctuation is much smaller
than for random states. Therefore we have mainly studied the short-time
dynamic process from such an ordered state to calculate $z$ and $\eta$.
On the other hand it is also desirable to measure these exponents from the
$m_0=0$ initial states \cite{Li95}. In our work we now take the evolution for
the second moment of magnetization $M^2(t)$ from the $m_0=0$ to verify the
results of $\eta$ and $z$. By input of values of $\eta$ and $z$ already
obtained from the indices $c_1$ and $c_u$, we ckeck if they also satify the
relation $y=(d-\eta)/z$ in Eq.(\ref{index_y}). As expected we find they
are consistent with less than 2\% error for $L\geq 96$.

\begin{figure}[htbp!]\centering
\caption{Time evolution of magnetization for different initial
magnetizations $m_0$, plotted on a log-log scale. The curves of
$m_0=0.01$ and $m_0=0.00$ show nearly complete overlap.
      }
\includegraphics*[width=7.30cm]{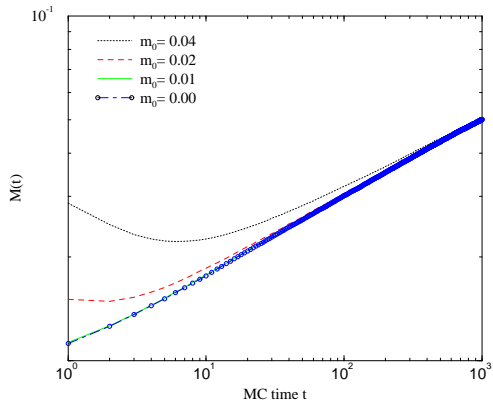}
\label{fig5}
\end{figure}
\section{\bf Conclusions}

We have made a first attempt to apply the STD approach to the 2D RBPF with
a continuous distribution of quenched disorders. Our simulations
verified that a second order phase transition has been induced.
Then the universal dynamic scaling behavior in the short-time dynamics is
used to estimate the exponents $\theta$ and $z$, as well the magnetic
critical exponent $\eta$. It is found that they really violate the Ising-like
universality class where $\theta=0.191(1)$ and $\eta=0.240(15)$ \cite{Li95}.
Our result of the dynamical exponent $z=3.41(6)$ is much bigger than that
for systems without disorder, and it causes the microscopic time $t_{mic}$
to be extended up to one hundred for the Binder cumulant from the ordered
initial state. The value for the exponent $\eta$ nearly completely
overlaps with those listed in Table 1, except that it is somewhat
smaller than that given in Ref.\cite{Olson99}. More accurate calculation
remains to be done to solve this controversy.

\section{\bf acknowledgements}
We acknowledge stimulating discussions with Prof. L. Sch\"ulke.
H.P.Y. would like to thank the Heinrich-Hertz-Stiftung for a fellowship
and be grateful to the hospitality of the University of Siegen where
the MC simulations were performed.


\end{document}